\documentclass{PoS}

\usepackage{amssymb,amsmath}
\usepackage{url}
\usepackage{bm}

\def\ie{i.e.\ }
\def\eg{e.g.\ }
\newcommand{\Recola}{{\sc RECOLA}}
\newcommand{\RecolaTwo}{{\sc RECOLA2}}
\newcommand{\Reptil}{{\sc REPT1L}}
\newcommand{\FORM}{{\sc FORM}}
\newcommand{\Feynrules}{{\sc Feynrules}}
\newcommand{\Sarah}{{\sc Sarah}}
\newcommand{\UFO}{{\sc UFO}}

\title{Recola2: a one-loop matrix-element generator for BSM theories and SM effective field theory}

\ShortTitle{Recola2}

\author{Ansgar Denner, \speaker{Jean-Nicolas Lang}\\
        Julius-Maximilians Universit\"at W\"urzburg, 97074
        W\"urzburg, Germany\\
        E-mail: \email{jlang@physik.uni-wuerzburg.de}
        }

\author{Sandro Uccirati\\
        Universit\`a di Torino e INFN, 10125 Torino, Italy\\
        E-mail: \email{uccirati@to.infn.it}}

\abstract{
We present the \RecolaTwo{} library for the efficient generation and computation of
one-loop amplitudes in Beyond-Standard-Model theories. \RecolaTwo{} is based on
RECOLA, an efficient one-loop amplitude generator for the Standard Model, and
REPT1L, a newly developed tool to generate one-loop model files for RECOLA2 in a
fully automated way.
\RecolaTwo{} is able to operate with non-trivial extensions of the SM, \eg extended
Higgs sectors and effective field theories. We discuss first applications to
extended Higgs sectors and their renormalization.
}

\FullConference{13th International Symposium on Radiative Corrections\\
                24-29 September, 2017\\
                St. Gilgen, Austria}

\begin{document}

\section{Introduction}

One of the burning questions in particle physics concerns the precise nature of the
Higgs boson. In the Standard Model (SM) the Higgs mechanism is implemented in a
minimal fashion, predicting the existence of a single physical Higgs boson, but
many extensions of the SM require this Higgs boson to be accompanied by additional
scalar bosons in order to account for astrophysical observations.
To be compatible with current measurements the additional Higgs bosons can only
couple weakly to other SM particles, in particular to the gauge bosons. This
constraint requires any extended Higgs sector to be (almost) aligned with the
SM, giving rise to SM-like tree-level couplings between SM particles.
Thus, interesting signatures are expected to arise only in the electroweak (EW)
corrections of the corresponding models, and these are needed to be computed in order
to distinguish a Beyond-Standard-Model (BSM) signature from the SM ``background''.
The so-called aligned and decoupling scenarios are not restricted to extended
Higgs sectors, and similar limits exist in theories with
an extended gauge group or extended space-time symmetry. In order to study all these
interesting extensions automated tools are required which allow for precision
predictions with realistic final states including QCD and EW corrections.



\section{Recola2: A general one-loop amplitude provider}

The QCD NLO revolution a decade ago was only the beginning of a continuous
improvement and development of new highly efficient algorithms for one-loop amplitudes.
Today the computation of QCD and EW one-loop amplitudes in the SM is fully
automated in various approaches, overcoming the deficits of the traditional
Feynman-diagram approach with sophisticated posterior algebraic simplification
or recursion techniques
\cite{Frixione:2015zaa,Cullen:2014yla,Chiesa:2015mya,Hahn:2000kx,Hahn:2016ebn}
and new recursive algorithms without or only partially
resorting to Feynman diagrams
\cite{Alwall:2014hca,Kallweit:2014xda,Buccioni:2017yxi,Actis:2012qn,Actis:2016mpe}.
With \RecolaTwo{} we use current technology and push limitations to the next
level, enabling the fully automated computation of tree and one-loop amplitudes
in BSM and SM effective field theories. 
As the successor of the \Recola{} library, \RecolaTwo{} is naturally based on
the very same off-shell recursion relations \cite{Actis:2012qn} without referring to Feynman
diagrams at any stage in the computation.
At tree-level it uses Berends-Giele off-shell recursion \cite{Berends:1987me} to
compute tree-level amplitudes directly. 
At one-loop level, amplitudes $\mathcal{M}_1$ are first decomposed
in terms of tensor coefficients $c$ and tensor integrals $T$ as
\begin{align}
  \mathcal{M}_1 = \sum_k c_{k,\mu_1\ldots} T_k^{\mu_1\ldots}.
  \label{eq:loopamp}
\end{align}
The tensor coefficients, encoding the model and process dependence, are computed
fully recursively by means of an extended version of the off-shell algorithm by van
Hameren \cite{vanHameren:2009vq}, whereas the (model-independent) results for
tensor integrals are obtained from the Collier library
\cite{Denner:2014gla,Denner:2016kdg}. Even though being based on the same
algorithmic structure \RecolaTwo{} overcomes many shortcomings of the original
\Recola{} library.  In summary, \RecolaTwo{} has been generalized with respect
to \Recola{} in the following ways \cite{Denner:2017vms}:
\begin{itemize}
  \item All hard-coded structures, in particular SM-specific pieces of the code, have
    been replaced with dynamic structures, and any necessary model-dependent
    information is retrieved from \RecolaTwo-specific model files.

  \item The process-generation phase has been generalized to arbitrary
    processes involving new fields and higher $n$-point vertices.
    The algorithmic implementation has been improved by excessive use of
    recursive functions.  A new feature of
    \RecolaTwo{} allows to write the recursion relations as \FORM{}
    code \cite{Ruijl:2017dtg}, permitting the reconstruction of tree and one-loop
    amplitudes analytically. This feature is predominantly used in the
    renormalization and computation of rational terms for new models (see
    Section \ref{se:ren}).  Finally, the process generation has been extended to deal with
    model files formulated in the Background-Field Method (BFM).

  \item The process-computation phase has been extended accordingly to deal with
    new couplings, Lorentz structures and higher $n$-point vertices.  Most
    notably, new composite Lorentz structures, \eg linked to non-renormalizable
    operators, are supported dynamically, making computations involving higher
    powers of loop momenta in Feynman rules possible. For instance, this
    ingredient has been used for the implementation of the $R_\xi$-gauge at
    one-loop level.

  \item So far \RecolaTwo{} is restricted to theories with scalars, Dirac
    fermions and vector bosons. An enhanced version with the support of Majorana
    fermions will be developed in the future.
\end{itemize}

\section{Framework for one-loop renormalization}\label{se:ren}

In \RecolaTwo{} the computation of renormalized one-loop amplitudes consists of the
computation of the loop amplitude \eqref{eq:loopamp}, in addition to rational
terms of type $R_2$ and counterterms:
\begin{align}
  \mathcal{M}_1^{Ren} = \sum_c c_{\mu_1\ldots} T^{\mu_1\ldots}_c +
  \mathcal{M}_{\mathrm{R}_2} + \mathcal{M}_{\mathrm{CT}}.
  \label{eq:loopampren}
\end{align}
The appearance of the extra rational terms is due to the use of dimensional
regularization and the fact that we compute the tensor coefficients in exactly 4
dimensions.
These remnants, emerging from the $4-D$ dimensional part of tensor coefficients
and the pole part of tensor integrals, can, however, be implemented as special
Feynman rules treated on equal footing with the counterterms of the theory.
Thus, one step towards the evaluation of \eqref{eq:loopampren} for generic
theories is to compute the rational terms of type $\mathrm{R}_2$ and the
counterterms for each model file once and for all. The second step in our
approach is to bring the Feynman rules into the right format suited for our
recursion techniques, resulting in \RecolaTwo-specific model files. In
summary, our approach towards the automated derivation of one-loop model files
consists of the following 3 steps:
\begin{itemize}
  \item[1)] In the first step we define the model in terms of a Lagrangian with a
    subsequent derivation of the tree-level Feynman rules in the \UFO{} format
    \cite{Degrande:2011ua}.  We use tools like \Feynrules{} 
    \cite{Christensen:2008py,Alloul:2013bka} and \Sarah{} \cite{Staub:2013tta} to
    perform this task.
  \item[2)] Then, we compute the rational and counterterm vertices
    analytically using the newly developed tool \Reptil{}
    \cite{Denner:2017vms}. The main ingredient for this kind of computations is
    the ability of \RecolaTwo{} to derive one-loop vertex functions analytically.
    This is done using the \RecolaTwo{} process generator and \FORM{} for the analytic
    reconstruction of amplitudes from the process skeletons.
    For more details on the computation and renormalization procedure we refer
    to the original reference \cite{Denner:2017vms}.
  \item[3)] Finally, all the (additional) Feynman rules can be mapped to the
    recursion kernels, resulting in a renormalized model file for \RecolaTwo{}
    ready for computations of, in principle, arbitrary processes in the
    underlying theory.
\end{itemize}

\section{Application to extended Higgs sectors}

As a first application for \Reptil{} and \RecolaTwo{} we have chosen the
Higgs-singlet extension of the SM and the Two-Higgs-Doublet model.
The renormalization is performed as far as possible following the
on-shell renormalization of the SM, \ie all
parameters associated to fields are renormalized in the on-shell or Complex-Mass
scheme \cite{Denner:1999gp,Denner:2005fg,Denner:2006ic}.
For the gauge couplings we support the conventional schemes:
The strong coupling constant can be renormalized in a fixed and variable flavour
(running) scheme, whereas the electroweak coupling constant can be renormalized in the
Thomson limit, at the Z~pole, or in the Fermi scheme.
The remaining independent parameters can be identified as mixing angles, which,
at leading-order, describe the mixing between fields in the interaction and mass
eigenbases.
As their renormalization is controversial, we support various schemes
encountered in the literature:
\begin{description}
  \item [$\overline{\boldsymbol{\mathrm{MS}}}$:] We support a large class of
    $\overline{\mathrm{MS}}$ schemes which are all defined 
      by requiring a vanishing pole part of a suited vertex function.
      The actual vertex function is irrelevant as long as the parameter
      to be renormalized appears in the corresponding counterterm vertex.
      The parameter of choice can be either a mixing angle or a parameter of the
      Higgs potential.\footnote{Since the number of
      independent parameters is fixed, this step requires to trade
      one of the mixing angles for one of the Higgs self couplings $\lambda_i$.}
      In addition, we support the renormalization in different tadpole
      counterterm schemes, which, however, leads to gauge-dependent results in
      combination with an $\overline{\mathrm{MS}}$ renormalization of the mixing
      angles unless the Fleischer--Jegerlehner tadpole scheme \cite{Fleischer:1980ub,Krause:2016oke,Denner:2016etu} is used.
      In this way the different $\overline{\mathrm{MS}}$ schemes proposed in 
      Refs.~\cite{Denner:2016etu,Freitas:2002um,Altenkamp:2017ldc} are obtained.

    \item [on-shell:] The on-shell schemes are defined by requiring that certain
      mixing energies\footnote{For the considered models the condition is
      imposed on the neutral Higgs-boson mixing and the
      charged/pseudoscalar Higgs-to-Goldstone-boson mixing.} vanish at
    specific phase-space points.  The procedure is 
      gauge dependent and requires an ad hoc prescription to render the
      $S$-matrix gauge independent. This procedure amounts to, roughly speaking,
      fixing the value for the mixing angle counterterm in a particular gauge.
      We implemented the schemes proposed in
      Refs.~\cite{Krause:2016oke,Espinosa:2002cd,Bojarski:2015kra} using the BFM for
      the quantum gauge-parameter set to one and translated the results to the
      conventional formulation in the 't~Hooft--Feynman gauge.
\end{description}

The model files together with the \RecolaTwo{} library are publicly
available under:
\begin{center}
\url{recola.hepforge.org}
\end{center}
Each of the renormalized model files has also been derived in the
Background-Field formulation of quantum field theory. This serves as a
powerful complementary computation method, enabling cross checks of
all the ingredients entering one-loop renormalized amplitudes
\eqref{eq:loopampren}.

\section{Summary}
We presented the \RecolaTwo{} library for the automated generation and
computation of tree and one-loop amplitudes in Beyond-Standard-model and
effective field theories.  The \RecolaTwo{} library requires renormalized model
files in a special format which are derived in a fully automated way using the
newly developed tool \Reptil{} from nothing but the Feynman rules in the \UFO{}
format.  We demonstrated the capabilities using the example of extended Higgs
sectors.  The code \RecolaTwo{} and a set of corresponding model files
are publicly available under \url{recola.hepforge.org}.

\end{document}